\documentclass[a4paper,11pt]{article}
\author{Luca Degiovanni\footnote{Dipartimento di Matematica, Universit\`a di Torino, Italy.}}
\date{}
\title{A note on the superintegrability of the Toda lattice\footnote{Work supported by the PRIN research project \textit{``Geometric methods in the theory of nonlinear waves and their applications''} of the italian MIUR.}}

\usepackage{latexsym}
\usepackage{amssymb}

\newcommand{\BE}{\begin{equation}}
\newcommand{\EE}{\end{equation}}
\newcommand{\BAS}{\begin{eqnarray*}}
\newcommand{\EAS}{\end{eqnarray*}}

\newcommand{\D}{\mathrm{d}}
\newcommand{\E}{\mathrm{e\,}}
\newcommand{\Uno}{\mathrm{1\kern-2pt{I}}}
\newcommand{\Mezzo}{\frac{1}{2}}

\newcommand{\Pois}[2]{\{#1,#2\}}
\newcommand{\Lie}[2]{\lbrack #1,#2 \rbrack}

\newtheorem{Prop}{Proposition}
\newtheorem{Cor}{Corollary}
\newtheorem{Oss}{Remark}

\newenvironment{Dim}
{\par\medskip\upshape\noindent\textbf{Proof}:}
{\hfill$\Box$\bigskip}

\begin{document}
\maketitle
\abstract{\noindent The superintegrability of the non-periodic Toda lattice is explained in the framework of systems written in action-angles coordinates. Moreover, a simpler form of the first integrals is given.}

\section{Background results}
The non-periodic Toda lattice is a celebrated completely integrable system formed by $n$ 
particles (with coordinates $x_i$ and momenta $p_i$) moving on a line according to the Hamilton equations
\BE\label{HamEq}
\begin{array}{rclcrcl}
\dot{x}_1 &=& p_1 && \dot{p}_1 &=& -\E^{x_1-x_2} \\
\dot{x}_2 &=& p_2 && \dot{p}_2 &=& \E^{x_1-x_2} - \E^{x_2-x_3} \\
&\vdots& && &\vdots \\
\dot{x}_i &=& p_i && \dot{p}_i &=& \E^{x_{i-1}-x_i} - \E^{x_i-x_{i+1}} \\
&\vdots& && &\vdots \\
\dot{x}_n &=& p_n && \dot{p}_n &=& \E^{x_{n-1}-x_n} \\
\end{array}
\EE
the Hamiltonian of the system is
$$
H(q_1,..,q_n,p_1,..,p_n) = \frac{1}{2}\sum_{i=1}^{n}p_i^2 + \sum_{i=1}^{n-1} \E^{q_i- q_{i+1}}\,.
$$

Introducing the Lax pair
\BE\label{Lax1}
L=\Mezzo\left(
\begin{array}{ccccc}
-p_1&\E^{\frac{x_1-x_2}{2}}&0&\cdots&0  \\
\E^{\frac{x_1-x_2}{2}}&-p_2&\ddots&\ddots&\vdots  \\
0&\ddots&\ddots&\ddots&0  \\
\vdots&\ddots&\ddots&\ddots&\E^{\frac{x_{n\!-\!1}-x_n}{2}} \\
0&\cdots&0&\E^{\frac{x_{n\!-\!1}-x_n}{2}} &-p_n
\end{array}
\right)
\EE
$$
B=\Mezzo\left(
\begin{array}{ccccc}
0&-\E^{\frac{x_1-x_2}{2}}&0&\cdots&0  \\
\E^{\frac{x_1-x_2}{2}}&0&\ddots&\ddots&\vdots  \\
0&\ddots&\ddots&\ddots&0  \\
\vdots&\ddots&\ddots&\ddots&-\E^{\frac{x_{n\!-\!1}-x_n}{2}} \\
0&\cdots&0&\E^{\frac{x_{n\!-\!1}-x_n}{2}} &0
\end{array}
\right)
$$
the Hamilton equations~(\ref{HamEq}) imply the Lax equation $\dot{L}=\Lie{L}{B}$, thus the $n$ eigenvalues $\lambda_i$ of $L$ are first integrals. Since one can show that the functions $\lambda_i$ are independent and mutually in involution, the Toda lattice is completely integrable. Other equivalent sets of first integrals in involution are given by the trace of the first $n$ powers of the matrix $L$ and by the coefficients of its characteristic polynomial $\det(L-\lambda\Uno)$.

In the paper~\cite{ADS2005}, a set of $n-1$ new independent constants of motions are found, obviously not in involution with the functions  $\lambda_i$. The basic step is to construct the so called \emph{Moser coordinates} $(r_i,\lambda_i)$ introduced in~\cite{Mos1975b} using the \emph{Weyl function}
$$
f(\lambda)=\frac{\Delta_{n-1}}{\Delta_n}\,,
$$
where $\Delta_k$ is the subdeterminant obtained by cancelling the last $n-k$ rows and columns of the matrix $(\lambda\Uno-L)$. The Weyl function $f(\lambda)$ has $n$ simple poles (coinciding with the eigenvalues $\lambda_i$ of $L$) with positive residua such that $\sum_i\mathrm{res}_{\lambda_i} f(\lambda)=1$. The positive functions $r_i$ are related to the residua through the formula
\BE\label{fraz_semplice}
f(\lambda)=\frac{1}{K^2}\,\sum_{i=1}^n\frac{r_i^2}{\lambda-\lambda_i}\,, \quad K=\E^{\Mezzo x_n}
\EE
implying
\BE\label{def_r}
r_i^2=K^2\mathrm{res}_{\lambda_i} f(\lambda)\,.
\EE
The map from coordinates $(x_i,p_i)$ to coordinates $(\lambda_i,r_i)$, with $\lambda_1<\lambda_2<\cdots<\lambda_n$ and $r_i>0$, is one to one and can be formally inverted. Indeed, $\sum_i\mathrm{res}_{\lambda_i} f(\lambda)=1$ implies $K^2=\sum_i r_i^2$ and $f(\lambda)$ admits a continued fraction expansion that goes back to Stieltjes~\cite{GK} and plays a crucial role in~\cite{Mos1975b}:
\BE\label{fraz_continua}
f(\lambda)=\frac{2}{2\lambda+p_n-\frac{\E^{x_{n-1}-x_n}}{2\lambda+p_{n-1}-\frac{\E^{x_{n-2}-x_{n-1}}}{
{\ddots \atop \quad2\lambda+p_2-\frac{\E^{x_1-x_2}}{2\lambda+p_1}}
}}}\,.
\EE
By comparing the expressions~(\ref{fraz_continua}) and~(\ref{fraz_semplice}) one finds that $p_i$ and $\E^{x_i}$ are rational functions of the coordinates $(\lambda_i,r_i)$.

Moser coordinates greatly simplify equations~(\ref{HamEq}): in fact, by differentiating the expression~(\ref{def_r}) and using the identity
$$
-\Mezzo\mathrm{res}_{\lambda_i} \dot{f}(\lambda)=(\lambda_i+\Mezzo p_n)\,\mathrm{res}_{\lambda_i} f(\lambda)
$$
proved in~\cite{Mos1975b}, one obtains
\BAS
r_i\dot{r}_i &=& \frac{\dot{K}}{K}r_i^2+\Mezzo K^2\,\mathrm{res}_{\lambda_i} \dot{f}(\lambda)\\
&=& \left[\frac{\dot{K}}{K}-\Mezzo p_n-\lambda_i\right]r_i^2\,.
\EAS
The equations~(\ref{HamEq}), written in Moser coordinates, then become
\BE\label{MosEq}
\left\{
\begin{array}{rcl}
\dot{\lambda}_i&=&0 \\
\dot{r}_i&=&-r_i\lambda_i
\end{array}
\right.
\EE
because $\dot{K}=\Mezzo Kp_n$ and the eigenvalues $\lambda_i$ are constant of the motion.

In the paper~\cite{ADS2005} it is shown, by direct computation, that for $k=1,\ldots,n-1$ the functions
\BE\label{DamInt}
H_k=\left(\frac{r_k}{r_{k+1}}\right)^2\exp\;\left(2\frac{\lambda_k-\lambda_{k+1}}{\displaystyle{\sum_{i=1}^n\lambda_i}}\sum_{i=1}^n\ln r_i\right)
\EE
are first integrals, and moreover the function $H_k$ are independent both mutually and with respect to the $\lambda_i$. Hence the Toda lattice is a superintegrable system.

\section{A set of simpler first integrals}
The functions $H_k$ given by~(\ref{DamInt}) are pretty complicated. Furthermore, the fact that they are first integrals depends only on the very special form~(\ref{MosEq}) of the system's equations. Indeed, as already noted in~\cite{ADS2005}, by setting $\rho_i=\ln r_i$ these equations become
\BE\label{ActAngEq}
\left\{
\begin{array}{rcl}
\dot{\lambda}_i&=&0 \\
\dot{\rho}_i&=&-\lambda_i
\end{array}
\right.
\EE
and it holds $\Pois{\lambda_i}{\lambda_j}=\Pois{\rho_i}{\rho_j}=0$, $\Pois{\lambda_i}{\rho_j}=\delta_{ij}$, i.e. the coordinates $(\rho_i,\lambda_i)$ can be regarded as action-angles coordinates for the Toda lattice.

It is then very natural to look for a simpler form for the constants of motion, and try to generalize the argument to any system in action-angle coordinates. Using the coordinates $(\rho_i,\lambda_i)$ one gets
\BAS
H_k&=&\exp2\,(\rho_k-\rho_{k+1})\exp2\,\left(\frac{\lambda_k-\lambda_{k+1}}{\displaystyle{\sum_{i=1}^n\lambda_i}}\sum_{i=1}^n\rho_i\right)\\
&=& \exp2\;\frac{(\rho_k-\rho_{k+1})\displaystyle{\sum_{i=1}^n\lambda_i}+(\lambda_k-\lambda_{k+1})\displaystyle{\sum_{i=1}^n\rho_i}}{\displaystyle{\sum_{i=1}^n\lambda_i}}
\EAS
and therefore, since $\displaystyle{\sum_{i=1}^n\lambda_i}$ is again a first integral, also the functions
\BE\label{NewInt}
\widetilde{H}_k=(\rho_k-\rho_{k+1})\displaystyle{\sum_{i=1}^n\lambda_i}+(\lambda_k-\lambda_{k+1})\displaystyle{\sum_{i=1}^n\rho_i}
\EE
are constants of motion.

This is a particular case of a general situation:
\begin{Prop}
If the phase space of an Hamiltonian system admits a set of coordinates $(\theta_i,I_i)$ such that the Hamilton equations take the form
\BE\label{ActAngEq2}
\left\{
\begin{array}{rcl}
\dot{I}_i&=&0 \\
\dot{\theta}_i&=&\omega_i(I_k)
\end{array}
\right.
\EE
then  either the functions $I_k$ or the functions 
$$
K_k=\theta_k\sum_{i=1}^n\omega_i-\omega_k\sum_{i=1}^n\theta_i
$$
are first integrals, for $k=1,\ldots,n$.

Moreover if an index $k_\ast$ exists, such that in an open dense subset of the phase space $\omega_{k_\ast}\neq0$ and $\sum_i\omega_i\neq0$, then the number of functions $K_k$ functionally independent between themselves and with the functions $I_k$ is exactly $2n-1$.
\begin{Dim}
Differentiating and using the Hamilton equations one obtains
$$
\dot{K}_k=\omega_k\sum_{i=1}^n\omega_i-\omega_k\sum_{i=1}^n\omega_i=0\,.
$$
At most $2n-1$ of the function $K_k$ are independent because
$$
\sum_{k=1}^nK_k=\sum_{k=1}^n\theta_k\sum_{i=1}^n\omega_i-\sum_{k=1}^n\omega_k\sum_{i=1}^n\theta_i=0
$$
One can always suppose that the non-vanishing $\omega_{k_\ast}$ is $\omega_n$ then, setting
\BAS
\eta &=& \D I_1\wedge\ldots\wedge\D I_n \\
\D\Theta &=& \sum_{i=1}^n\D\theta_i\\
\Omega &=& \sum_{i=1}^n\omega_i
\EAS
because $\eta\wedge\D\omega_j=0$ one has
\BAS
&&\hspace{-20pt}\eta\wedge\D K_2\wedge\ldots\wedge\D K_n = \eta\wedge(\Omega\D\theta_2-\omega_2\D\Theta)\wedge\ldots\wedge(\Omega\D\theta_n-\omega_n\D\Theta)\\
&&\hspace{-20pt}=\Omega^{n-2}\eta\wedge\big[\Omega\D\theta_2\wedge\ldots\wedge\D\theta_n -\\
&&\hspace{-10pt}\omega_2\D\Theta\wedge\D\theta_3\wedge\ldots\wedge\D\theta_n-\ldots-\omega_n\D\theta_2\wedge\ldots\wedge\D\theta_{n-1}\wedge\D\Theta\big]
\EAS
Hence the coefficient of $\eta\wedge\D\theta_1\wedge\ldots\wedge\D\theta_{n-1}$ in $\eta\wedge\D K_2\wedge\ldots\wedge\D K_n$ is $(-1)^{n-1}\Omega^{n-2}\omega_n$; this means that in the open dense subset of the phase space, where $\Omega\neq0$ and $\omega_n\neq0$, the differentials of the function $K_2\ldots K_n$ and $I_k$ are linearly independent, and therefore the corresponding functions are functionally independent.
\end{Dim}
\end{Prop}
\begin{Cor}
If the Hamilton equations are given by~(\ref{ActAngEq}), then the functions $K_k-K_{k+1}$ take the form~(\ref{NewInt}) and hence they are $2n-1$ constants of motion functionally independent between themselves and with the functions $\lambda_i$.
\end{Cor}
\begin{Oss}
Another equivalent set of first integrals for the Hamilton equations~(\ref{ActAngEq2}) are the ``generalized angular momenta''
$$
J_{ij}=\theta_i\omega_j-\omega_i\theta_j
$$
that are simpler although their functional dependence is more cumbersome.
\end{Oss}

\section{Final remarks}
The superintegrability of the non-periodic Toda lattice is a special case of the superintegrability of a wider class of Hamiltonian systems. The key point is the existence in the phase space of the non-periodic Toda lattice of a global set of coordinates (the well known Moser coordinates~\cite{Mos1975b}) of action-angles type. The extra constants of the motion found in~\cite{ADS2005} for the non-periodic Toda lattice are indeed functional combinations of simpler first integrals, that are defined for all systems admitting global action-angles coordinates.


\begin{thebibliography}{0}
\bibitem{ADS2005} \textsc{M.\ Agrotis, P.\ A.\ Damianou, C.\ Sophocleus}, \textit{The Toda lattice is super-integrable}, math-ph/0507051.
\bibitem{GK} \textsc{F.\ R.\ Gantmacher, M.\ G.\ Krein}, \textit{Oscillation matrices and kernels and small vibrations of mechanical systems}, revised edition, AMS Chelsea Publishing 2002.
\bibitem{Mos1975b}\textsc{J.\ Moser}, \textit{Finitely many mass points on the line under the influence of an exponential potential --- An integrable system},  Lect.\ Notes Phys.\  \textbf{38} (1976), 97--101.

\end{thebibliography}
\end{document}